\documentclass[preprint,pra,onecolumn]{revtex4}%
\usepackage{amsfonts}
\usepackage{amsmath}
\usepackage{amssymb}
\usepackage{graphicx}
\usepackage{caption2}%
\providecommand{\U}[1]{\protect \rule{.1in}{.1in}}
\renewcommand \caption{}

\begin{document}
\title{Robust quantum cryptography with a heralded single-photon source based on the
decoy-state method}
\author{Qin Wang$^{1,2}$}
\email{qinw@kth.se}
\author{Wei Chen$^{1}$}
\author{Guilherme Xavier$^{2}$}
\author{Marcin Swillo$^{2}$}
\author{Tao Zhang$^{1}$}
\author{Sebastien Sauge$^{2}$}
\author{Maria Tengner$^{2}$}
\author{Zheng-Fu Han$^{1}$}
\author{Guang-Can Guo$^{1}$}
\author{Anders Karlsson$^{2}$}
\affiliation{$^{1}$Department of Physics, Key Laboratory of Quantum Information, CAS, USTC,
230026, Hefei, China}
\affiliation{$^{2}$Department of Microelectronics and Applied Physics, Royal Institute of
Technology (KTH), Electrum 229, SE-164 40 Kista, Sweden}

\begin{abstract}
In this paper, we describe a robust quantum cryptography scheme with a
heralded single photon source based on the decoy-state method, which has been
shown by numerical simulations to be advantageous compared with many other
practical schemes not only with respect to the secure key generation rate but
also to secure transmission distance. We have experimentally tested this
scheme, and the results support the conclusions from numerical simulations
well. Although there still exist many deficiencies in our present systems,
it's still sufficient to demonstrate the advantages of the scheme. Besides,
even when cost and technological feasibility are taken into account, our
scheme is still quite promising in the implementation of tomorrow's quantum cryptography.

PACS number(s): 03.67.Dd, 42.65.Yj, 03.67.Hk

\end{abstract}
\maketitle

\section{\bigskip Introduction}

Cryptography plays an important role in the field of communication, the goal
of it is to render messages between legitimate users (usually called Alice and
Bob) but incomprehensible to Eve (a malicious eavesdropper). However,
classical cryptography is based on conjectured computational complexity, and
its security is thus threatened by the advancement in mathematical algorithms
and computational power. Compared with the classical method, quantum
cryptography has unprecedented advantages because it does not based on
computational complexity, and its unconditional security is ensured by the
\textquotedblleft battle tested\textquotedblright \ theory of quantum mechanics
\cite{bb84,eker,shor,maye1,lo1}.

Since the first protocol of quantum cryptography was put forward by Bennett
and Brassard in 1984 \cite{bb84}, (hereafter called BB84 protocol,) quantum
cryptography has\ been widely investigated and developed by large numbers of
researchers and scientists, not only in theory, but also by experimental
implementations. Unfortunately, there always exists some discrepancies between
the \textquotedblleft in principle\textquotedblright \ unconditional security
and realistic systems. Therefore, people have to take practical usability into
account when estimating a quantum cryptosystem, just as summarized in Ref.
\cite{gisin}:%

\begin{align}
\text{Infinite security}  &  \Rightarrow \text{ Infinite cost}\nonumber \\
&  \Rightarrow \text{ Zero practical interest.}%
\end{align}

In current practice, an attenuated laser (\textit{i.e., }emitting a weak
coherent state WCS) or a parametric down-conversion source (PDCS) are employed
in most quantum cryptosystems. A WCS is quite easy to implement. However, it
contains a large vacuum state probability amplitude and unneglectable
multiphoton probability amplitude when attenuated to a single photon level,
which is fatal to photon-number-splitting (PNS) attack
\cite{hutt,bras,lutk0,lutk1}\ and some other attacks. To compensate the
security, one has to attenuate a laser into a quite low intensity
(\textit{e.g.} 0.1 photon/ pulse), resulting in a low secure key generation
rate and a limited transmission distance.

Fortunately, the so-called decoy-state method was given out
\cite{hwang,wang1,lo2}. The main idea of the decoy-state method is to randomly
mix extra decoy transmission events with the true signal transmission events.
The decoy transmission events and the signal transmission events have the same
characteristics (such as wavelength, bandwidth and timing information, etc.)
except for different intensities. Therefore, for a given random photon state,
Eve is unable to judge whether it comes from a decoy transmission event or
from a signal transmission event. Hence she has to perform the same operations
on them. On the other hand, the legitimate users, Alice and Bob, could
estimate the behavior of vacuum, single-photon and multi-photon states
individually just by doing some counting measurements and classical
communications. As a result, Eve's eavesdropping with be detected. It has been
shown that the decoy-state method has significantly improved the performance
of a quantum cryptography with practical systems.

The heralded single-photon source (HSPS) from parametric down-converted (PDC)
processes has also been widely investigated in recent years \cite{maria, ried,
mori,alib,trif}. Its sub-Poissonian photon number distribution makes it
suitable for the implementation of quantum cryptography. (It has already been
proven that a sub-Poissonian distributed source is superior to a Poissonian
one in the quantum cryptography \cite{subp}.)

Combining the advantages of the decoy state method and the HSPS, we have
proposed some schemes that applies both of them in a quantum cryptography
setup \cite{qin1,qin2,qin3,qin4}. In Ref. \cite{qin1,qin2,qin3}, only
theoretical aspects are discussed (for simplicity, those models in them all
assumed some ideal conditions, \textit{i.e.} the idler and the signal photons
in different paths have the same photon distributions before being detected.
However, they can be quite different in practice because of existing coupling
loss and some other factors.). In Ref. \cite{qin4}, some preliminary
experimental results are presented. Here in this paper, we will describe our
theory and experiment in detail and include some experimental improvements as well.

This paper is organized as follows: At first, in section II, we will introduce
some basic theory on HSPS and report some experimental results from our group;
In section III, we will introduce our scheme in detail, and also do some
numerical simulations to show the advantages of our scheme compared with other
practical schemes; In section IV, we will describe our experimental setup and
experimental processes, and compare our experimental results with theoretical
predictions; Subsequently, we will discuss some deficiencies existing in our
present systems, and then make suggestions on how to improve them in section
V; Finally, the conclusions are drawn and the future prospects are spelled out.

\section{Theory and experiment in HSPS}

In recent years, a HSPS based on a PDC process has been investigated by many
groups. With improvements of down conversion in waveguides and four-wave
mixing, a lot of encouraging results have been reported \cite{maria, ried,
mori,alib,trif}.

The main idea of the HSPS is to use one photon (heralding) of a photon pair to
announce the arrival of the other one (heralded). The temporal statistics of a
HSPS can be controlled by utilizing a \emph{prior information }(\textit{i.e.,}
original distribution) extracted from the photon pairs. As said in Ref.
\cite{maria}, during the nondegenarate spontaneous parametric down conversion
(SPDC) process, if a pulsed pump laser is used, as long as the coherence time
of the emission, $\Delta t_{c}$, is much longer than the duration of the pump
pulse, $\Delta t$, \textit{i.e.,} $\Delta t_{c}\gg$ $\Delta t$, (in practice
easily obtained by using ultrafast (fs) pulse pump lasers), a single emission
process will take place, giving an thermal photon number distribution. In
contrast, when a continuous wave (CW) laser is used, as long as $\Delta t_{c}$
is much shorter than the gating period of the detector, a large number of
independent SPDC processes will be present, each thermally distributed, but
collectively resulting in a Poisson distribution. However, the
\textquotedblleft original\textquotedblright \ distribution can be altered by
conditional gating. By choosing proper gating time and using an appropriate
correlation rate, a sub-Poissonian distributed HSPS can be obtained as the
result of postselections.

To quantify our source, let us introduce the second-order auto-correlation
function at zero-time delay:%
\begin{equation}
g^{(2)}(0)=\frac{2P_{m\geq2}}{P_{m\geq1}^{2}}. \label{a}%
\end{equation}
(It is known that, the value of $g^{(2)}(0)$ could be used to classify a
source between a Poisson ($g^{(2)}(0)=1$), a sub-Poisson ($g^{(2)}(0)<1$) and
a super-Poisson ($g^{(2)}(0)>1$) distribution.) $P_{m\geq k}$ is the
probability to find at least $k$ photons within a gating period, which can be
expressed as:
\begin{equation}
P_{m\geq k}=P^{cor}P_{m\geq k-1}^{acc}+(1-P^{cor})P_{m\geq k}^{acc}, \label{b}%
\end{equation}
where $P^{cor}$ is the correlation rate of photon pairs, \textit{i.e.} the
probability that we can predict the existence of a heralded photon when a
heralding one was detected. (Its value equals unity under perfect experimental
conditions, \textit{i.e., }when there is no coupling loss, no transmission
loss, and no detection loss etc.) $P_{m\geq k}^{acc}$ is the probability that
at least $k$ accidental photons are present within a gating period, which
comes from the \textquotedblleft original\textquotedblright \ statistical
distribution of the down-converted light (the coherence time of the single
photons ($\Delta t_{c}\sim10$ ps) is much less than the integration time
($2.5$ ns), as a result, those signal photons which are not coming from the
same SPDC process as the heralding one (it means they are not truly correlated
photon pairs) may also contribute to the final coincidence counts.). It is
given by:%
\begin{equation}
P_{m\geq k}^{acc}=1-%
{\displaystyle \sum \limits_{i=0}^{k-1}}
\frac{\mu^{i}}{i!}e^{-\mu},\text{ \  \  \ }(k\geq2) \label{c}%
\end{equation}
where $\mu$ is the average photon number per gating time (before detection),
$\mu=R_{s}\cdot \Delta t_{gate}$, $R_{s}$ is the mean photon number per second
(before detection), and $\Delta t_{gate}$ is the gating time of the detector.

Moreover, the probability of getting exactly $n$ photons within the gating
time is:%
\begin{equation}
P(n)=P_{m\geq n}-P_{m\geq n+1},\text{ \  \ }(n\geq2). \label{d}%
\end{equation}

To be noted, the vacuum state probability should be treated independently,
which can be stated as:%
\begin{equation}
P(0)=P^{cor}\cdot d_{i}+(1-P^{cor})\cdot e^{-\mu}, \label{e}%
\end{equation}
where $d_{i}$ is the dark count probability of the detector for heralding photons.

Consequently, the single-photon probability is:%
\begin{equation}
P(1)=1-P(0)-P_{m\geq2}. \label{f}%
\end{equation}

Considering Eqs. (\ref{a})-(\ref{f}), in order to analyze a HSPS, we should at
first get to know the values of $P^{cor}$ and $\mu$. In the following, we will
explain how their values can be determined in an experiment.

\begin{figure}[ptb]
\begin{center}
\includegraphics[scale=0.7]{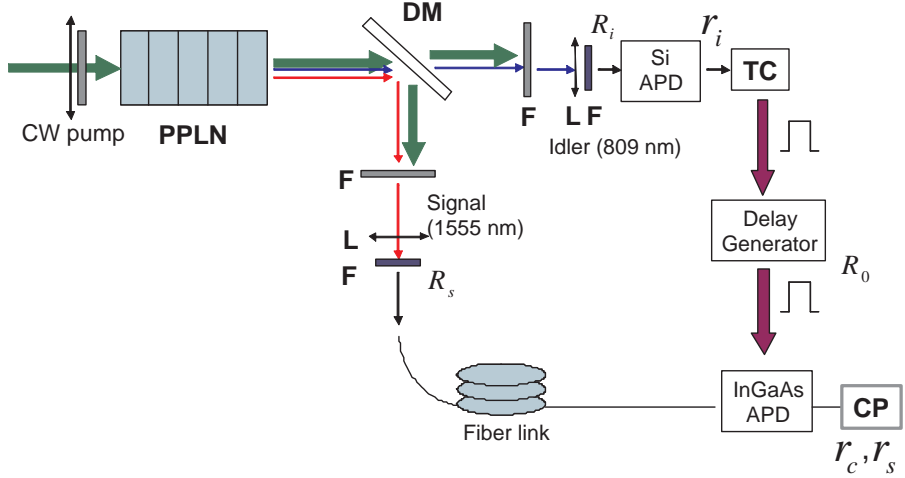}
\end{center}
\caption{Fig. 1: (Color online) Schematics of the experimental setup to
produce a HSPS. PPLN: periodically-poled LiNbO$_{3}$ crystal; DM: dichroic
mirror; F: filter; L: lens; TC: time chopper; CP: counter processing; $R_{i}$
and $R_{s}$ are the photon rates inside the fibers for idler and signal
respectively; $r_{i}$ is the single photon counting rate at the Si-detector;
$R_{0}$ is the triggering rate; $r_{c}$ and $r_{s}$ are the counting rates at
the InGaAs detector triggered by idler photons and triggered by random pulses
individually.}%
\label{Fig1}%
\end{figure}

Our experimental setup for the HSPS is shown in Fig. 1. We use a CW laser at
the wavelength of $532$ nm to pump a periodically-poled LiNbO$_{3}$ (PPLN)
crystal of $50$ mm length to generate non-degenerate correlated photon pairs.
The photon at the wavelength of $809$ nm is called idler, and the one at the
wavelength of $1555$ nm is called signal. After separation by a dichroic
mirror (DM), they are coupled into different detectors. For the idler photons,
we used a Si-based APD (PerkinElmer SPCM-AQR-14) with a detection efficiency
of about $50\%$; for the signal photons, a InGaAs-APD (id200-SMF) operating in
gated Geiger mode is used. Whenever there is an idler photon being detected,
an electronical pulse is sent out. After the time chopper (TC, the details
about it will be described in section IV), each electronical pulse is used to
trigger the InGaAs detector (with a gating time of $2.5$ ns), then the
coincidence count, $r_{c}$, will be obtained; $r_{i}$ is the Si-detector
single counting rate; $r_{s}$ is the InGaAs detector single counting rate with
random gating (the gating frequency is $R_{0}$), to provide the mean
accidental photon number; $R_{0}$ is the heralding rate, whose value can be
different from $r_{i},$ because of the dead/delay time of the pulse generator
used; Besides, $\eta_{i}$ ($\eta_{s}$) and $d_{i}$ ($d_{s}$) are the detection
efficiency and dark count probability of idler (signal) photons respectively;
$R_{i}$ ($R_{s}$) is the corresponding photon number for idler (signal) before detection.

When the InGaAs detector is randomly gated, the detection probability can be
written as:%
\begin{equation}
P_{\det}^{Ran}\equiv \frac{r_{s}}{R_{0}}=1-(1-P^{acc})(1-P_{dark}),
\end{equation}
where $P^{acc}$ ($=1-e^{-\eta_{s}R_{s}\Delta t_{gate}}$) and $P_{dark}$
($=\frac{d_{s}}{R_{0}}$) are the corresponding probabilities caused by
accidental photons and dark counts.

From these relations, it can be deduced that:%
\begin{equation}
R_{s}=\frac{1}{\eta_{s}\Delta t_{gate}}\ln \frac{R_{0}-d_{s}}{R_{0}-r_{s}}.
\label{rs}%
\end{equation}
When the InGaAs detector is gated by idler photons, the detection probability
is:%
\begin{equation}
P_{\det}\equiv \frac{r_{c}}{R_{0}}=1-(1-P^{cor})(1-P^{acc})(1-P_{dark}),
\end{equation}
because the detection events can be caused by correlated photons, accidental
photons or dark counts. The equations above lead to:%
\begin{equation}
P^{cor}=1-\frac{R_{0}-r_{c}}{R_{0}-d_{s}}e^{\eta_{s}R_{s}\Delta t_{gate}}.
\label{pcor}%
\end{equation}

Obviously, all the parameters in Eqs. (\ref{rs}) and (\ref{pcor}) can be
experimentally measured, giving the values of $R_{s}$ and $P^{cor}$. By
substituting them into Eqs. (2)-(7), we can finally calculate the photon
number distribution of our source shown in Table I.

\begin{figure}[ptb]
\caption{\textbf{Table I. }The measured photon-number distributions of our
HSPS under different triggering frequencies.}%
\newline \includegraphics[scale=0.8]{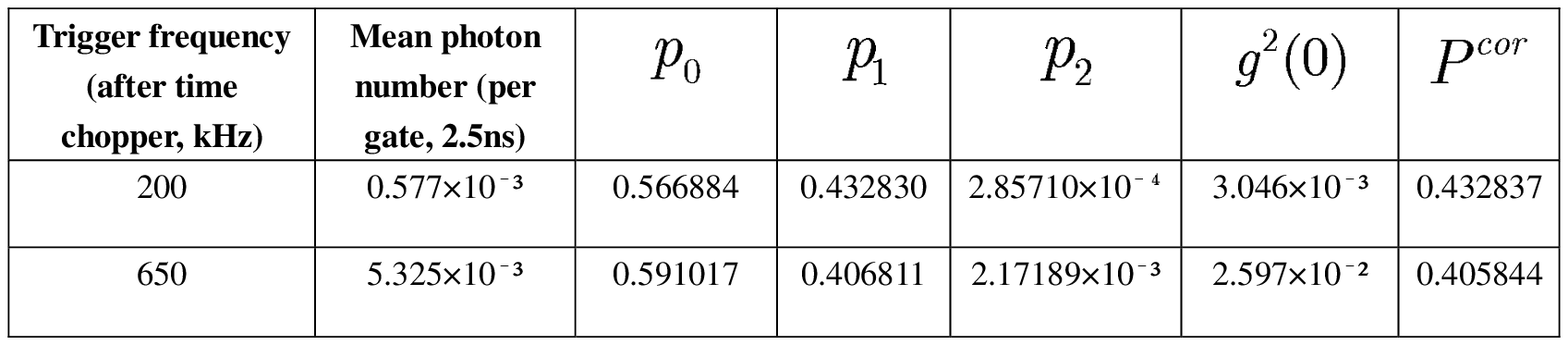}\newline \end{figure}

It can be seen from Table I that, because of a CW laser being used, the
multi-photon probability of our source has substantially been depressed. By
carefully optimizing the alignment of our optical systems, we can get a
sub-Poissonain distributed HSPS with the single photon probability of about
$40\%$ (it's about $30\%$ in \cite{qin4}), which substantially improves the
performance of a quantum cryptography as shown in Fig. 2.

\section{Quantum cryptography with a HSPS based decoy-state method}

Similarly to Ref. \cite{qin1,qin2,qin3}, we have used a three-intensity
($\mu^{\prime}$, $\mu$, $\mu_{0}$) decoy-state method, where $\mu^{\prime}$,
$\mu$ and $\mu_{0}$ are the mean photon number per gate for the signal light,
the decoy light and the \textquotedblleft vacuum\textquotedblright%
\ light\ individually. The counting probabilities for $\mu^{\prime}$ and $\mu$
can be written as:%
\begin{align}
Q_{\mu^{\prime}}  &  =%
{\displaystyle \sum_{i=0}^{\infty}}
Y_{n}P_{\mu^{\prime}}(n),\label{q1}\\
Q_{\mu}  &  =%
{\displaystyle \sum_{i=0}^{\infty}}
Y_{n}P_{\mu}(n), \label{q2}%
\end{align}
Defining $Y_{n}$ to be the yield of an n-photon state, \textit{i.e.}, the
conditional probability of a detection event at Bob's side given that Alice
sends out an n-photon state, which is essentially a sum of two contributions,
background and true signal, \textit{i.e.,} $Y_{n}=Y_{0}+1-(1-\eta)^{n}$.
$\eta$ is the combined detection efficiency and transmittance between Alice
and Bob. $Y_{0}$ is Bob's background rate, which includes the detector dark
count and other background contributions such as the stray light from timing
pulses. The gain, $G_{n}$ is the product of the probability of Alice sending
out an n-photon state and the conditional probability of Alice's n-photon
state, which is given by: $G_{n}=Y_{n}P(n)$. $P_{\mu^{\prime}}(n)$ ($P_{\mu
}(n)$) is the n-photon number probability in the source of $\mu^{\prime}$
($\mu$).

Furthermore, the average quantum bit error ratios (QBER) for $\mu^{\prime}$
and $\mu$ are given by:%
\begin{align}
E_{\mu^{\prime}}  &  =\frac{%
{\displaystyle \sum_{i=0}^{\infty}}
Y_{n}P_{\mu^{\prime}}(n)e_{n}}{Q_{\mu^{\prime}}},\\
E_{\mu}  &  =\frac{%
{\displaystyle \sum_{i=0}^{\infty}}
Y_{n}P_{\mu}(n)e_{n}}{Q_{\mu}},
\end{align}
where $e_{n}$ is the quantum bit error probability of an n-photon state.

Eq. (\ref{q1}) and (\ref{q2}) lead to:%
\begin{align}
&  P_{\mu^{\prime}}(2)Q_{\mu}-P_{\mu}(2)Q_{\mu^{\prime}}\nonumber \\
&  =Y_{0}[P_{\mu^{\prime}}(2)P_{\mu}(0)-P_{\mu}(2)P_{\mu^{\prime}}%
(0)]+Y_{1}[P_{\mu^{\prime}}(2)P_{\mu}(1)-P_{\mu}(2)P_{\mu^{\prime}%
}(1)]\nonumber \\
&  +%
{\displaystyle \sum_{i=2}^{\infty}}
Y_{n}[P_{\mu^{\prime}}(2)P_{\mu}(n)-P_{\mu}(2)P_{\mu^{\prime}}(n)].
\end{align}
Considering Eqs. (\ref{a})-(\ref{f}) and the values of $P^{cor}$ for $\mu$ and
$\mu^{\prime}$ in Table I, one can show that for our setup $%
{\displaystyle \sum_{i=2}^{\infty}}
Y_{n}(P_{\mu^{\prime}}(2)P_{\mu}(n)-P_{\mu}(2)P_{\mu^{\prime}}(n))\leq0$,
which leads to:%
\begin{align*}
Y_{1}  &  \geq \frac{P_{\mu^{\prime}}(2)Q_{\mu}-P_{\mu}(2)Q_{\mu^{\prime}%
}-Y_{0}(P_{\mu^{\prime}}(2)P_{\mu}(0)-P_{\mu}(2)P_{\mu^{\prime}}(0))}%
{(P_{\mu^{\prime}}(2)P_{\mu}(1)-P_{\mu}(2)P_{\mu^{\prime}}(1))},\\
e_{1}  &  \leq \frac{E_{\mu^{\prime}}Q_{\mu^{\prime}}-e_{0}Y_{0}P_{\mu^{\prime
}}(0)}{Y_{1}p_{1}^{\prime}(\mu^{\prime})}.
\end{align*}
When taking statistical fluctuations into account, we can get a lower bound
for $Y_{1}$ and an upper bound for $e_{1}$:%
\begin{align}
Y_{1}^{L}  &  =\frac{P_{\mu^{\prime}}(2)Q_{\mu}^{L}-P_{\mu}(2)Q_{\mu^{\prime}%
}^{U}-Y_{0}^{U}(P_{\mu^{\prime}}(2)P_{\mu}(0)-P_{\mu}(2)P_{\mu^{\prime}}%
(0))}{(P_{\mu^{\prime}}(2)P_{\mu}(1)-P_{\mu}(2)P_{\mu^{\prime}}(1))},\\
e_{1}^{U}  &  =\frac{E_{\mu^{\prime}}Q_{\mu^{\prime}}^{U}-e_{0}Y_{0}^{L}%
P_{\mu^{\prime}}(0)}{Y_{1}^{L}p_{1}^{\prime}(\mu^{\prime})}, \label{e1u}%
\end{align}
where $Q_{\mu}^{L}\equiv Q_{\mu}\left(  1-\frac{10}{\sqrt{N_{\mu}Q_{\mu}}%
}\right)  $, $Q_{\mu^{\prime}}^{U}\equiv Q_{\mu^{\prime}}\left(  1+\frac
{10}{\sqrt{N_{\mu^{\prime}}Q_{\mu^{\prime}}}}\right)  $, $Q_{\mu^{\prime}%
}E_{\mu^{\prime}}^{U}\equiv Q_{\mu^{\prime}}E_{\mu^{\prime}}\left(
1+\frac{10}{\sqrt{N_{\mu^{\prime}}Q_{\mu^{\prime}}E_{\mu^{\prime}}}}\right)
$, $Y_{0}^{L}\equiv \left(  1-\frac{10}{\sqrt{N_{0}Y_{0}}}\right)  $, and
$Y_{0}^{U}\equiv \left(  1+\frac{10}{\sqrt{N_{0}Y_{0}}}\right)  $ \cite{10sqrt}.

After error correction and privacy amplification, we can get the final key
generation rate from the signal ($\mu^{\prime}$) \cite{GLLP,koas}:%

\begin{equation}
R\geq q\left \{  -Q_{\mu^{\prime}}f\left(  E_{\mu^{\prime}}\right)
H_{2}\left(  E_{\mu^{\prime}}\right)  +G_{0}+G_{1}^{L}\left[  1-H_{2}\left(
e_{1}^{U}\right)  \right]  \right \}  ,
\end{equation}
where the factor $q$ ($=\frac{1}{2}$) comes from the cost of basis mismatch in
the Bennett-Brassard 1984 (BB84) protocol, (it's $\frac{1}{4}$ when a
one-detector scheme is used), $f(E_{\mu^{\prime}})$ is a factor represents the
cost of error correction given existing error correction systems in practice.
We use $f(E_{\mu^{\prime}})=1.22$ here \cite{bras}; In addition, $G_{0}$
$\equiv Y_{0}P_{\mu^{\prime}}(0)$; $G_{1}^{L}$ $\equiv Y_{1}^{L}P_{\mu
^{\prime}}(1)$; $H_{2}\left(  x\right)  $ is the binary Shannon information
function, given by
\[
H_{2}\left(  x\right)  =-x\log_{2}(x)-(1-x)\log_{2}(1-x).
\]

\bigskip We have used decoy states in the above deduction processes for
$G_{1}^{L}$ and $e_{1}^{U}$. However, we will consider the case without decoy
states in the following, in order to make a comparison. Without decoy state,
we have to do a pessimistic assumption in the estimation of $G_{1}^{L}$ and
$e_{1}^{U}$, \textit{i.e.,} we have to assume that the photons that fail to
arrive at Bob's side all come from single photon states. As a result, we get
the lower bound of $G_{1}$ as:%
\begin{equation}
G_{1}^{L}=Q_{\mu^{\prime}}-G_{0}-%
{\displaystyle \sum_{i=2}^{\infty}}
P_{\mu^{\prime}}(n).
\end{equation}
It's the same as Eq. (\ref{e1u}), corresponding upper bound value of $e_{1}$
can also be obtained.

Using the formulas above and considering different distributions, we can give
a comparison between our scheme using a HSPS based decoy-state method and
other practical schemes, including using a HSPS but without decoy-state
method, WCS with (or without) decoy-state method, and an ideal single photon
source. For the sake of fairness, during the comparison in all the schemes, we
use the BB84 protocol and assume the same experimental conditions,
\textit{i.e.,} the same dark count probability $0.8\times10^{-5}$/gate, the
same detection efficiency $7.5\%$, and the same misalignment of the system
$e_{\det ector}\sim2.5\%$. Corresponding numerical simulation results are
shown in Fig. 2. Clearly, compared with other practical schemes, our scheme
can tolerate the highest total loss, that also means the highest key
generation rate under fixed loss. Moreover, if a HSPS with $70$ percent single
photon probability (reported in \cite{SWP2007}) is used, its performance can
come close to the ideal single photon case.

\begin{figure}[ptb]
\begin{center}
\includegraphics[scale=0.8]{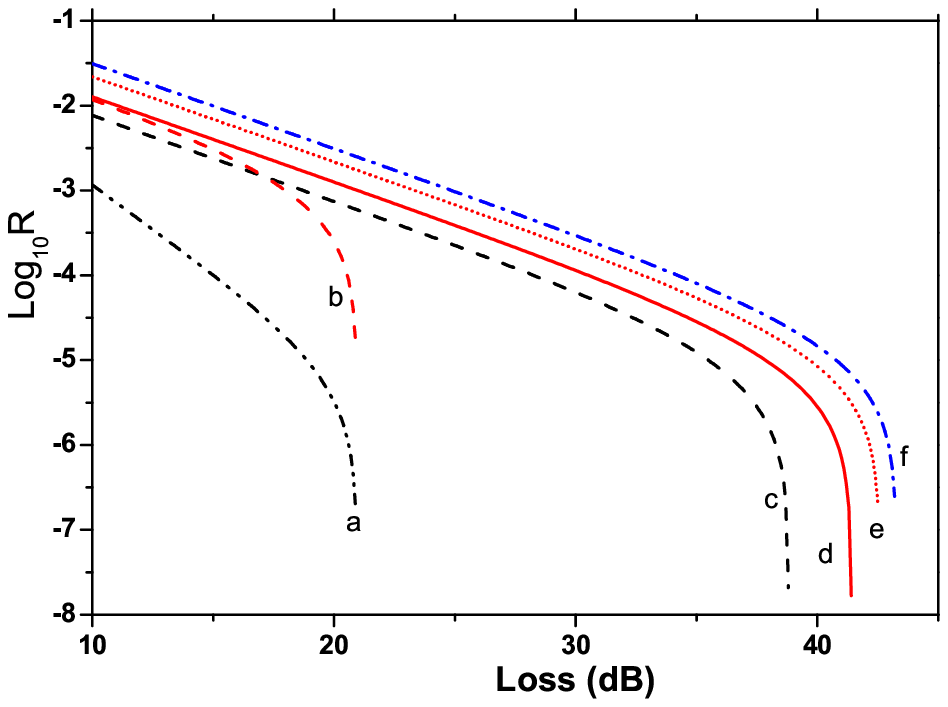}
\end{center}
\caption{Fig. 2: (Color online) The key generation ratio vs. the total losses
comparing several different schemes. The numerical simulations are done in the
case of: a) With WCS and without decoy-state method. b) With HSPS and without
decoy-state method; c). With WCS based decoy-state method (with optimal values
of $\mu^{\prime}$ at each point and an infinite number of decoy states). d)
With HSPS based decoy-state method with $P_{cor}$=40\% ($\mu^{\prime
}=5.325\times10^{-3}$ and $\mu=6.600\times10^{-4}$, these parameters come from
our experiment. After numerical simulation, we also find the key ratio is
stable with moderate variations of the value of $\mu^{\prime}$ or $\mu$). e)
With HSPS based decoy-state method with $P_{cor}$=70\% ($\mu^{\prime
}=5.325\times10^{-3}$ and $\mu=6.600\times10^{-4}$). f) With the ideal SPS.
(Noted: For fair comparison, we consider using infinite number of signal
pulses, so we don't take statistical fluctuation into account in all these
lines above.)}%
\label{Fig2}%
\end{figure}

\section{Experimental implementation}

\begin{figure}[ptb]
\begin{center}
\includegraphics[scale=0.9]{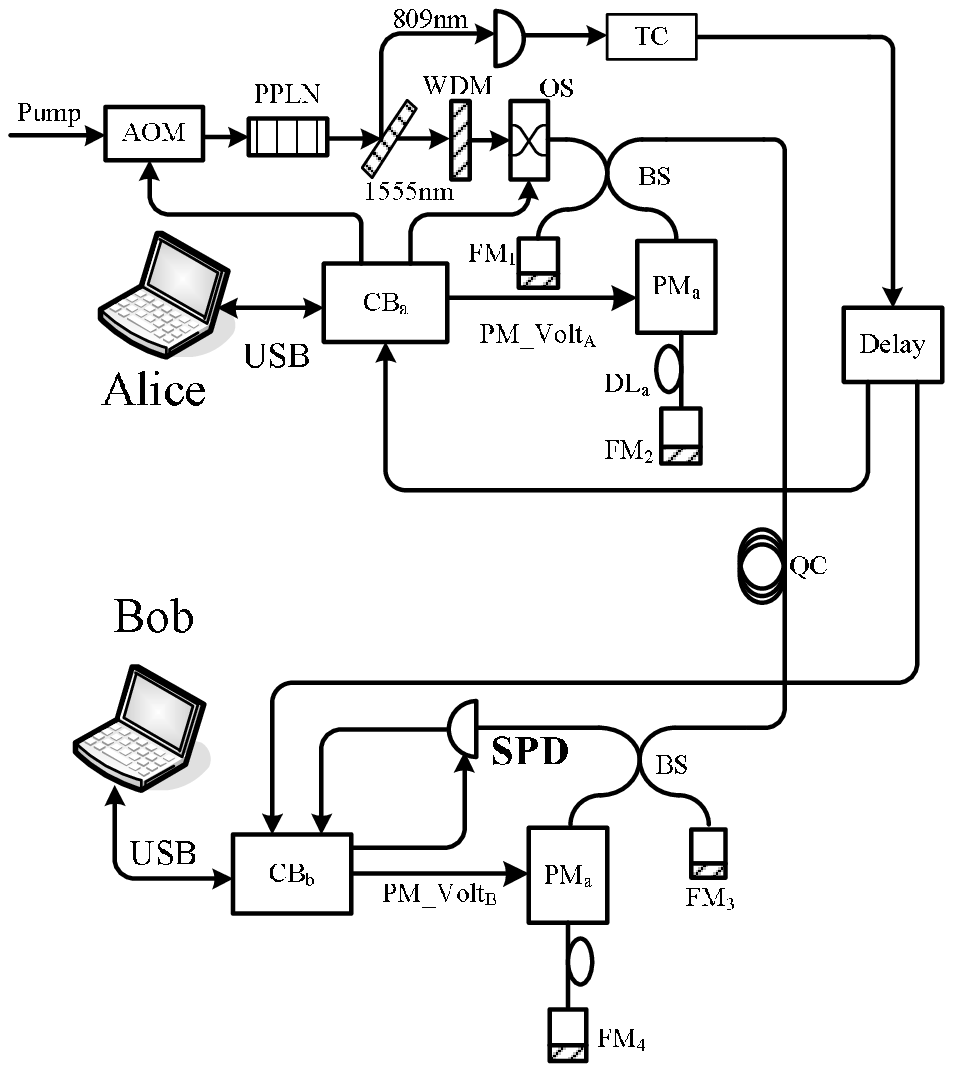}
\end{center}
\caption{Fig. 3: (Color online) The experimental setup of our quantum key
transmission system. PPLN: periodically-poled LiNbO$_{3}$; AOM:
acousto-optical-modulator; WDM: wavelength-division multiplexing; OS: optical
switch; TC: time chopper; BS: beam-splitter; FM: Faraday Mirror; PM: phase
modulator; DL: delay line; QC: quantum channel; SPD: single photon detector;
CB: control board. }%
\label{Fig3}%
\end{figure}

Our experimental setup is shown in Fig. 3. Using the same structures as in
Fig. 1, we can get a HSPS with narrow bandwidth ($0.8$ nm FWHM) and a single
photon number probability of about $40\%$ (see Table I). (The single photon
probability is improved compared with before reported in \cite{qin4}. Because
here we use a new pump laser whose coherence length and spectrum has a little
difference from the former one, which will inevitably influence the following
focusing and collecting processes. On the other hand, the single photon
probability obtained has close correlations with the final coincidence
counting rate, and the final coincidence counting rate is so sensitive to the
optical alignment and optical focusing settings. So we try to readjust the
positions of focusing lenses and re-optimize the optical alignment, which
result in an increased coincidence counting rate, and also an improved single
photon probability.) Then the heralded single photons are transmitted from
Alice to Bob through $25$ km of spooled SMF-28 fiber (attenuation: $0.2$
dB/km), incorporating a one-way Faraday-Michelson (F-M) cryptosystem
\cite{mo1}. We use a four-state \cite{US patent} and one-detector phase-coding
scheme, which is immune to time-shift attacks \cite{qi1,zhao1}, faked-state
attacks \cite{faked state}, Trojan horse attacks \cite{Trojan horse}, and can
also been proven to be secure against any other standard individual or
coherent attacks.

In order to generate down-converted light with three intensities ($\mu
^{\prime}$, $\mu$, $\mu_{0}$), on one hand we place an acousto-optic-modulator
(AOM, $3.5$ dB loss) in front of the PPLN crystal, on the other hand, we use a
fiber pig-tailed optical switch (OS, $0.6$ dB loss) at the arm of signal
photons ($1555$ nm). By controlling both of them in our program (changing
between $\mu^{\prime}$ and $\mu$ with AOM, and changing between $\mu$ and
$\mu_{0}$ with OS), we can randomly generate signals with three different mean
photon numbers: [$\mu^{\prime}$, $\mu$, $\mu_{0}$] = [$5.325\times10^{-3}$,
$0.660\times10^{-3}$, $0.577\times10^{-5}$]. (Because of an imperfect
isolation ratio of the optical switch ($\sim20$ dB), we don't produce a real
vacuum state, but generate a low mean photon number for $\mu_{0}$. We use its
counting rate instead of $Y_{0}$ for the estimation of $Y_{1}$, resulting in a
lower estimated value of $Y_{1}$. This also gives a lower key generation
rate.) The ratio of heralding pulses (\textit{i.e.,} gating instances) between
the three intensities is about $10:4:1$. In order to minimize the impact that
the power change would have on the triggering rate (since when we change the
pump intensity, we change the intensities of both $809$ nm and $1555$ nm
photons), we inserted a fixed dead time after each electrical pulse generated
from the Si-APD, calling it time chopper (TC). This way, when the power is
increased, the triggering rate does not increase in the same ratio, and
according to experimental verification, the dark count probability of our
InGaAs APD does not change significantly with the dead time implemented. This
dead time was implemented in the software controlling the whole QKD session,
and it was also important for our electronic circuit to be able to keep the
transmission synchronized. In addition, in order to get a higher visibility in
the F-M interferometers ($>95\%$, without removing any dark counts), we use a
wavelength-division multiplexing (WDM) filter to further narrow the bandwidth
of the signal photons to $0.4$ nm at FWHM. (The spectra measured before the
WDM and after the WDM of $1555$ nm photons are shown in Fig. 5(a). Fig. 5(b)
shows the interference curve of our F-M interferometer measured with a
"strong" light after the WDM filter.)

\begin{figure}[ptb]
\begin{center}
\includegraphics[scale=0.8]{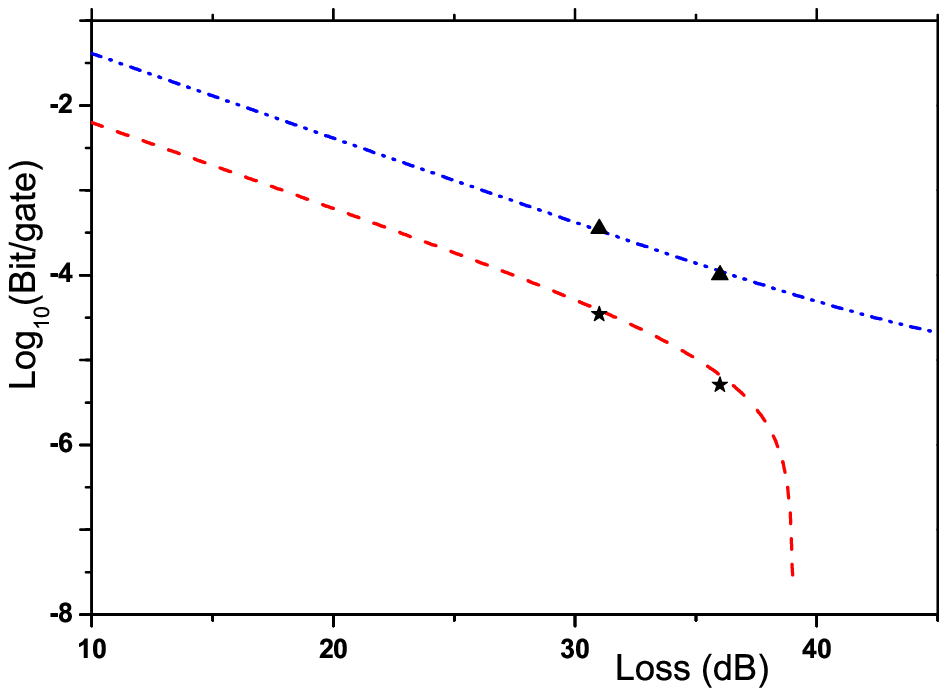}
\end{center}
\caption{Fig. 4: (Color online) Comparing the theoretical values and
experimental results in coincidence counting rate and the final secure key
rate. The top line represents the theoretical counting rate for signal photons
($\mu^{\prime}$); the bottom line represents the theoretical secure key rate
(taking statistical fluctuation into account). For each line, we investigated
two points at the total loss of 31dB and 36dB individually. The stars and
triangles are corresponding experimental results. }%
\label{Fig4}%
\end{figure}

\begin{figure}[ptb]
\begin{center}
\includegraphics[scale=0.8]{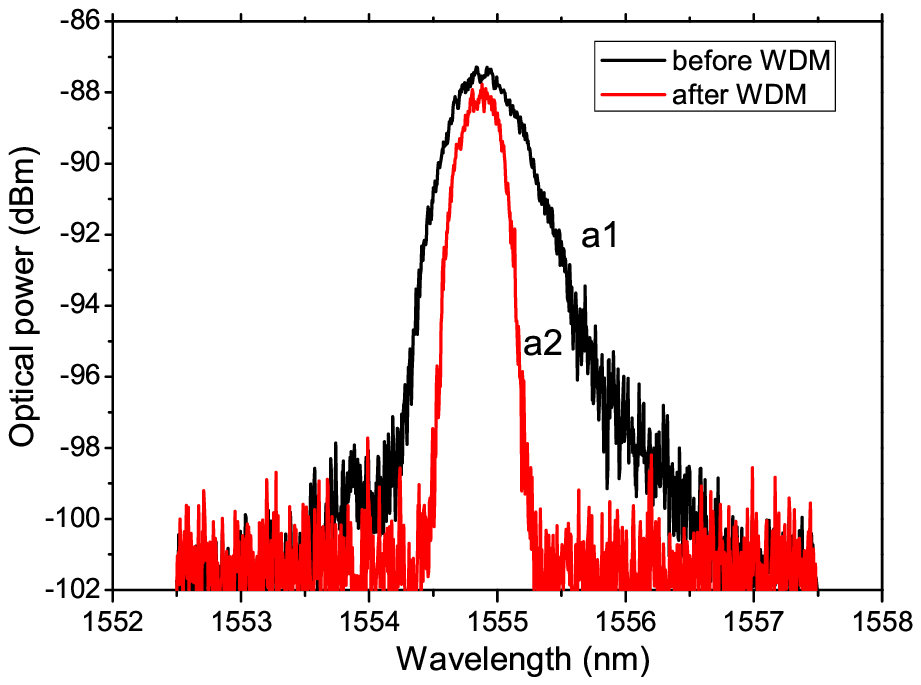}
\end{center}
\caption{Fig. 5(a): (Color online) Spectra of the signal photons (1555nm) with
a strong pump. a1) Measured before the WDM filter; a2) Measured after the WDM
filter.}%
\label{Fig5(a)}%
\end{figure}

\begin{figure}[ptb]
\begin{center}
\includegraphics[scale=0.8]{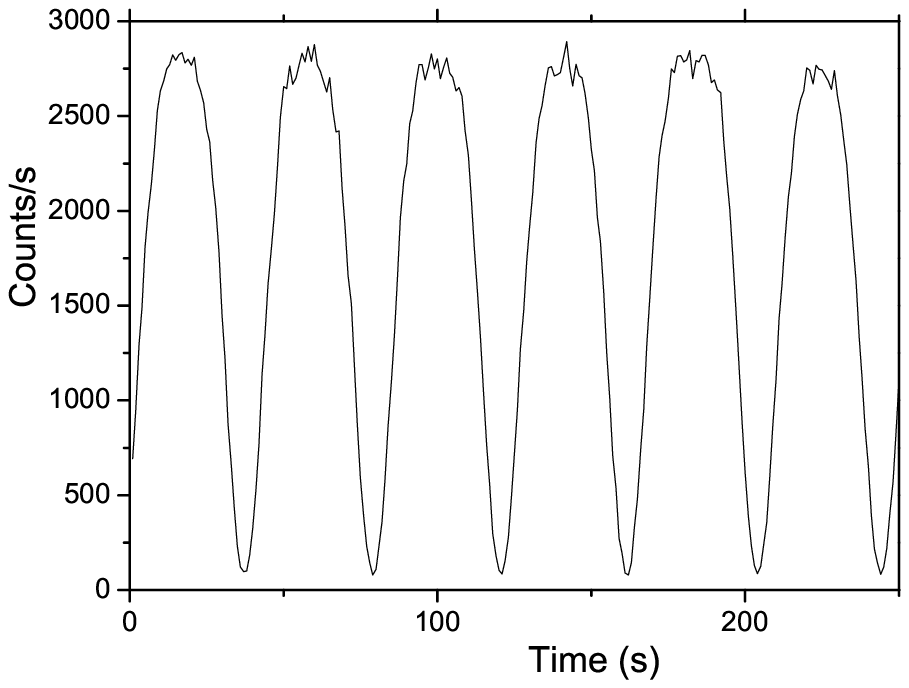}
\end{center}
\caption{Fig. 5(b): (Color online) The interference curve measured with a
"strong" light after the WDM filter passing through our F-M interferometer.
(Without removing any dark counts.)}%
\label{Fig5(b)}%
\end{figure}

In our quantum cryptosystem, in order to compensate for the phase-drift in the
interferometers caused by the environment, we adopted a scan and transmission
mode. The electronic circuit first generates an interference curve, to obtain
the correct working voltages for the phase modulators, and then quantum
transmission occurs. After a few blocks of data are exchanged, transmission is
stopped and scanning recommence to verify if the working point has changed for
the next transmission burst, then this pattern follows. For details we refer
to \cite{mo1}. The scan and transmission mode used makes the system quite
stable for several hours of continuous measurements. For example, during a
typical measurement of $12000$ s, (with effective transmission time about
$4200$ s, the scan and responding time are considerably longer than the
transmission time because of the low coincidence count rate), with a total of
$1.5$ $\times10^{9}$ triggering pulses, the detection efficiency is about
$7.5\%$, the "vacuum state" counting rate is about $0.8\times10^{-5}$/gate,
(we attribute $0.7\times10^{-5}$ coming from dark counts, and $0.1\times
10^{-5}$ coming from the leakage of the optical switch and the misalignment of
the system,) the counting rate, $Q_{\mu^{\prime}}$ ($Q_{\mu}$) and average
QBER, $E_{\mu^{\prime}}$ ($E_{\mu}$) are about $1.01\times10^{-4}$
($1.06\times10^{-4}$) and $6.33\%$ ($5.44\%$) respectively. After a total loss
of $36$ dB, we get a key generation rate of about $5.065\times10^{-6}$.
Finally, we obtain $5065$ secure key bits from a total of $143176$ coincidence
counts, which agrees well with the theoretical value as shown in Fig. 4 (using
a similar simulation model as \cite{qin1,qin3} in our theoretical predictions).

\section{Conclusions and prospects}

Our final key rate is lower than those in other systems reported
\cite{zhao2,rose,schm,peng,yuan0}, because there are substantial losses in our
present system. Apart from the insertion loss of the WDM filter, the optical
switch and a very low detection efficiency of our detector, the main loss
comes from the F-M interferometer used, because the signal photons have to go
through each phase modulator (PM) twice, and have to suffer losses from two
beam-splitters (BS). The aforementioned reasons caused a total loss of about
$31$ dB. However, using present technology, it's realistic to decrease the
loss to a lower value. For example, in order to remove the loss coming from
the WDM filter, a narrow bandwidth filter at the wavelength at $809$ nm can
used at heralding photons instead. This will not only avoid the loss of signal
photons, but also increase the correlation rate of photon pairs (it's $70\%$
reported in \cite{SWP2007}). Or if a cavity structure is used during the
parametric down-conversion processes, it will intrinsically depress the
bandwidth of the down-converted light. Moreover, to overcome the loss coming
from the F-M interferometer, a low loss Mach-Zehnder (M-Z) interferometer
(\cite{yuan0,yuan1}) can be used instead. Alternatively, if a
polarization-coding scheme is used to replace present phase-coding scheme, no
interferometer needed, so the scheme will suffer even less loss. In addition,
we can also use a better detector at $1555$ nm (with a lower dark count
probability ($\sim10^{-6}$) or a higher detection efficiency ($10-15\%$)) or
use a two-detector scheme. In all, it's quite realistic to reduce loss by
$15-18$ dB with present technology, corresponding to an increased transmission
distance of more than $100$ km.

In summary, though our present setup still contains many deficiencies, our
experimental results are sufficient to in principle demonstrate that our using
the HSPS based decoy state scheme could overcome many practical schemes in
loss tolerance, which also means it could give a highest key generation rate
under fixed loss. Besides, our scheme does not evoke higher costs or other
technological requirements than in any other schemes. Therefore, even when
practical usability is taken into account, it is still a very promising
candidate in the implementation of the quantum cryptography in the near
future. (The first three authors-Q. Wang, W. Chen and G. Xavier contribute
equally to the work.)

\bigskip

\  \  \  \  \  \  \  \  \  \  \  \  \  \  \  \  \  \  \  \  \  \  \  \  \  \  \  \  \  \  \  \  \  \  \  \  \  \  \  \  \  \textbf{Acknowledgement}%

Qin Wang is grateful to Prof. X. B. Wang (Tsinghua Univ.) and Dr. C. H. F.
Fung (Univ. of Toronto) for fruitful discussions, and Prof. G. Bj\"{o}rk for
valuable comments. This work was funded by the EU through the QAP (Qubit
Applications-015848) project, and the SECOQC project (FP6-2002-IST-1-506813),
the Swedish Science Research Council, the Swedish Foundation for Strategic
Research, the ECOC Foundation; and partly funded by the National Science
Foundation of China under Grant No. 60537020 and 60621064, Chinese Academy of
Sciences and International Partnership Project. G. B. Xavier thanks the
Brazilian agencies CAPES and CNPq for financial support.


\begin{thebibliography}{99}                                                                                               %


\bibitem {bb84}C. H. Bennett and G. Brassard, \emph{in Proceedings of IEEE
International Conference on Computers, Systems, and Signal Processing,
Bangalore, India} (IEEE, New York, 1984), p. 175.

\bibitem {eker}A. K. Ekert, Phys. Rev. Lett. \textbf{67}, 661 (1991).

\bibitem {shor}P. W. Shor and J. Preskill, Phys. Rev. Lett. \textbf{85}, 441 (2000).

\bibitem {maye1}D. Mayers, J. ACM \textbf{48}, 351 (2001).

\bibitem {lo1}H.-K. Lo and H.-F. Chau, Science \textbf{283}, 2050 (1999).

\bibitem {gisin}N. Gisin, G. Ribordy, W. Tittel, and H. Zbinden, Rev. Mod.
Phys. \textbf{74}, 145 (2002).

\bibitem {hutt}B. Huttner, N. Imoto, N. Gisin, and T. Mor, Phys. Rev. A
\textbf{51}, 1863 (1995); H. P. Yuen, Quantum Semiclassic. Opt. \textbf{8},
939 (1996).

\bibitem {bras}G. Brassard, N. L$\ddot{u}$tkenhaus, T. Mor, and B. Sanders,
Phys. Rev. Lett. \textbf{85}, 1330 (2000).

\bibitem {lutk0}N. L$\ddot{u}$tkenhaus and M. Jahma, New J. Phys. \textbf{4},
44 (2002).

\bibitem {lutk1}N. L$\ddot{u}$tkenhaus, Phys. Rev. A, \textbf{61}, 052304 (2000).

\bibitem {hwang}W. Y. Hwang, Phys. Rev. Lett. \textbf{91}, 057901 (2003).

\bibitem {wang1}X. B. Wang, Phys. Rev. Lett. \textbf{94}, 230503 (2005).

\bibitem {lo2}H. K. Lo, X. Ma, and K. Chen, Phys. Rev. Lett. \textbf{94},
230504 (2005).

\bibitem {maria}M. Tengner, and D. Ljunggren, e-print quant-ph/0706.2985v1.

\bibitem {ried}H. D. Riedmatten \textit{et al}., J. Mod. Opt. \textbf{51},
1637 2004.

\bibitem {mori}S. Mori, J. S\"{o}derholm, N. Namekata and S. Inoue, Opt.
Commun. \textbf{264}, 156 (2006).

\bibitem {alib}O. Alibart, D. B. Ostrowsky, P. Baldi, and S. Tanzilli, Opt.
Lett. \textbf{30}, 1539 (2005).

\bibitem {trif}A. Trifonov and A. Zavriyev, J. Opt. B: Quantum Semiclass. Opt.
\textbf{7,} S772 (2005).

\bibitem {subp}E. Waks, C. Santori, and Y. Yamamoto, Phys. Rev. \textbf{A}
\textbf{66}, 042315 (2002).

\bibitem {qin1}Q. Wang, X. B. Wang, and G. C. Guo, Phys. Rev. \textbf{A}
\textbf{75}, 012312 (2007).

\bibitem {qin2}Q. Wang, and A. Karlsson, Phys. Rev. \textbf{A 76,
}014309\textbf{ }(2007).

\bibitem {qin3}Q. Wang, X. B. Wang, G. Bj\"{o}rk and A. Karlsson, Europhys.
Lett. \textbf{79}, 40001 (2007).

\bibitem {qin4}Q. Wang \textit{et al., }Phys. Rev. Lett., \textbf{100}, 090501 (2008).

\bibitem {10sqrt}We estimated $e_{1}$ and $Y_{1}$ very conservatively as
within $10$ standard deviations, which promises a confidence interval for
statistical fluctuations of less than $1\times10^{-23}$.

\bibitem {GLLP}D. Gottesman, H.-K. Lo, N. L$\ddot{u}$tkenhaus, and J.
Preskill, Quantum Inf. Comput. \textbf{4}, 325 (2004).

\bibitem {koas}M. Koashi, e-print quant-ph/0609180v1.

\bibitem {SWP2007}A. Zavriyev and A. Trifonov, \emph{in Proceedings of single
photon workshop 2007} (Turin, Italy, 2007).

\bibitem {mo1}X. F. Mo, B. Zhu, Z. F. Han, Y. Z. Gui, and G. C. Guo, Opt.
Lett. \textbf{30}, 2632 (2005); Z. F. Han, X. F. Mo, Y. Z. Gui and G. C. Guo,
Appl. Phys. Lett. \textbf{86}, 221103 (2005).

\bibitem {US patent}M. J. LaGasse, Secure use of a single single-photon
detector in a QKD system, United States patent application 20050190922 (2005).

\bibitem {qi1}B. Qi, C. H. F. Fung, H. K. Lo, and X. F. Ma, Quantum Inf. Com.
\textbf{7}, 073 (2007)

\bibitem {zhao1}Y. Zhao, C. H. F. Fung, B. Qi, C. Chen, H. K. Lo, e-print quant-ph/0704.3253.

\bibitem {faked state}V. Makarov, J. Skaar, e-print quant-ph/0702262.

\bibitem {Trojan horse}N. Gisin, S. Fasel, B. Kraus, H. Zbinden, and G.
Ribordy, Phys. Rev. \textbf{A} \textbf{73}, 022320 (2006).

\bibitem {zhao2}Y. Zhao, B. Qi, X. F. Ma, H. K. Lo, and L. Qian, Phys. Rev.
Lett. \textbf{96}, 070502 (2006).

\bibitem {rose}D. Rosenberg \textit{et al}, Phys. Rev. Lett., \textbf{98},
010503 (2007).

\bibitem {schm}T. Schmitt-Manderbach \textit{et al}., Phys. Rev. Lett.,
\textbf{98}, 010504 (2007).

\bibitem {peng}C. Z. Peng \textit{et al}., Phys. Rev. Lett., \textbf{98},
010505 (2007).

\bibitem {yuan0}Z. L. Yuan, A. W. Sharpe, and A. J. Shields, Appl. Phys. Lett.
\textbf{90}, 011118 (2007).

\bibitem {yuan1}P. M. Intallura, M. B. Ward, O. Z. Karimov, Z. L. Yuan, P.
See, and A. J. Shields, e-print quant-ph/0710.0565.
\end{thebibliography}
\end{document}